# Discovery of Kiloparsec Extended Hard X-ray Continuum and Fe-Kα from the Compton Thick AGN ESO428–G014

G. Fabbiano[a], M. Elvis[a], A. Paggi[a], M. Karovska[a], W. P. Maksym[a], J. Raymond[a], G. Risaliti[b,c], Junfeng Wang[d]

a. Harvard-Smithsonian Center for Astrophysics, 60 Garden St. Cambridge MA 02138, USA
b. Dipartimento di Fisica e Astronomia, Università di Firenze, via G. Sansone 1, I-50019 Sesto Fiorentino (Firenze), Italy
c. INAF - Osservatorio Astrofisico di Arcetri, Largo E. Fermi 5, I-50125 Firenze, Italy
d. Department of Astronomy and Institute of Theoretical Physics and Astrophysics, Xiamen University, Xiamen, 361005, China

**ABSTRACT**

We report the discovery of kpc-scale diffuse emission in both the hard continuum (3-6 keV) and in the Fe Kα line in the Compton thick (CT) Seyfert galaxy ESO428-G014. This extended hard component contains at least ~24% of the observed 3-8 keV emission, and follows the direction of the extended optical line emission (ionization cone) and radio jet. The extended hard component has ~0.5% of the intrinsic 2-10 keV luminosity within the bi-cones. A uniform scattering medium of density 1 $cm^{-3}$ would produce this luminosity in a 1kpc path length in the bi-cones. Alternatively, higher column density molecular clouds in the disk of ESO428-G014 may be responsible for these components. The continuum may also be enhanced by the acceleration of charged particles in the radio jet. The steeper spectrum ($\Gamma \sim 1.7\pm0.4$) of the hard continuum outside of the central 1.5'' radius nuclear region suggests a contribution of scattered/fluorescent intrinsic Seyfert emission. Ultrafast nuclear outflows cannot explain the extended Fe Kα emission. This discovery suggests that we may need to revise the picture at the base of our interpretation of CT AGN spectra.

## 1. Introduction

Both the >3 keV continuum and the narrow, large Equivalent Width (EW) 6.4 keV Fe Kα emission of Compton Thick Active Galactic Nuclei (CT AGNs) are believed to originate from the interaction of hard photons from near the central black hole with a ~1-100 pc scale dense circumnuclear 'torus' (e.g. Matt et al. 1996, Gandhi et al 2015). This picture appeared to be confirmed by the high resolution *Chandra* imaging of the nearest CT AGN NGC4945 (D=3.7 pc; Marinucci et al 2012), where both hard continuum and Fe Kα line originate from a clumpy 8'' (~150 pc) diameter flattened structure, perpendicular to the main axis of the 'ionization cone'. But more recent results (Circinus, Arevalo et al 2014; NGC1068, Bauer et al 2015) show that both the hard continuum and Fe Kα emission may also have extended components on the direction of the ionization cone (~600 pc in Circinus, >140 pc in NGC1068).

In this letter we report the discovery of ~2 kiloparsec diameter extended hard continuum and Fe Kα emission associated with the CT AGN galaxy ESO428-G014.

ESO428-G014 (also called IRAS01745-2914, MCG-05-18-002), is a southern barred early-type spiral galaxy [SAB(r)], at a distance of ~23.3 Mpc (NED, Weyant et al., 2014; scale=112 pc/arcsec). This galaxy has a highly obscured (CT, $N_H>10^{25}$ cm$^{-2}$) Seyfert Type 2 nucleus, with a low ratio of hard X-ray (2-10 keV) observed flux to [OIII] λ5007 (Maiolino et al 1998). This source has the second highest [OIII] flux among CT AGNs after NGC1068 (Maiolino & Riecke 1995, Risaliti et al 1999), so it is expected to be the second intrinsically brightest CT AGN below 10 keV. This nucleus is associated with a one-sided curved 5''-long 6 and 20 cm. radio jet (Ulvestad & Wilson 1989) and with extended Hα and [OIII] optical line emission on a similar scale (Falcke et al 1996, 1998).

Based on the first ~30 ks *Chandra*/ACIS observation of ESO428-G014, Levenson et al (2006, ObsID 4866) found that the soft emission is dominated by an extended, line-dominated, photo-ionized X-ray component that also follows the extended optical line and radio emission, as in many nearby AGN (Wang et al, 2011a,b, 2014; Paggi et al 2012). At higher energies (>3 keV), the spectrum shows the hard, flat continuum and the prominent Fe Kα line (with EW=1.6 ± 0.5 keV), typical of CT AGNs. The cumulative *Chandra* exposure is now over 5 times longer than that used in Levenson et al (2006), allowing a detailed morphological study of the emission in different spectral bands.

## 2. Observations and Analysis

All the available *Chandra* ACIS-S observations were used in this paper, including our newly obtained 43 ks and 81 ks exposures (ObsIDs: 17030, 18745, 2016 January, P.I. Fabbiano), with a cumulative exposure time of 154.5 ks. We used CIAO 4.8 and CALDB 4.7.0 to reprocess and analyze the data. We inspected each observation for high background flares (≥ 3σ) and concluded that the entire data set was acceptable. Because this is a CT AGN, the X-ray luminosity of the nuclear source is heavily attenuated. Given its ACIS-S count rate, the point-like central source would only be ~1% piled-up[1] in ACIS-S, if point-like, but our analysis shows that the emission is extended. No 'cratering' is visible at the peak of the emission. The morphology of the extended emission and its similarity to that of the optical emission line region and radio emission (see below), also exclude instrumental effects. This allows us to explore the circumnuclear regions to the smallest radii allowed by the *Chandra* resolution.

We merged the three observations to produce a deeper data set for imaging analysis, following the CIAO threads[2], and using the longest observation (ObsID 18745) as the astrometric reference. All the data sets were processed to enable sub-pixel analysis (Tsunemi et al. 2001, Wang et al. 2011a). We first inspected visually each image in the 0.3-8 keV band and established that possible centroid shifts were below one ACIS instrument pixel (0.492"). We then used detected X-ray sources in each field to cross-match the images. Since a focus of this study is the analysis of the Fe Kα 6.4 keV line at high spatial resolution, we also created images in the 6.0-6.6 keV energy band, which is dominated by this line. This line image has the advantage of being dominated by the nuclear position in CT AGN. Examining the 6.0-6.6 keV images from each observation, we concluded that an additional shift, after applying the solution based on cross-matching other field sources, was needed for ObsID 17030. The final shifts (in units of native ACIS pixels) are Δx=0.036, Δy=0.0508 for ObsID 04866 and Δx=0.473, Δy=-0.1256 for ObsID 17030.

---

[1] Chandra Proposers' Observatory Guide, section 6.16,
cxc.harvard.edu/proposer/POG/html/chap6.html; see also Chandra ABC Guide to Pile Up,
http://cxc.harvard.edu/ciao/download/doc/pileup_abc.ps.
[2] URL: http://cxc.harvard.edu/ciao/threads/

Fig. 1 shows the 3-8 keV spectrum from the 1.5-8" annulus spectrum (red), centered on the centroid of the 0.3-8.0 keV emission, at RA = 7:16:31.208, Dec = -29:19:28.68, which is mostly due to the extended emission (see below).

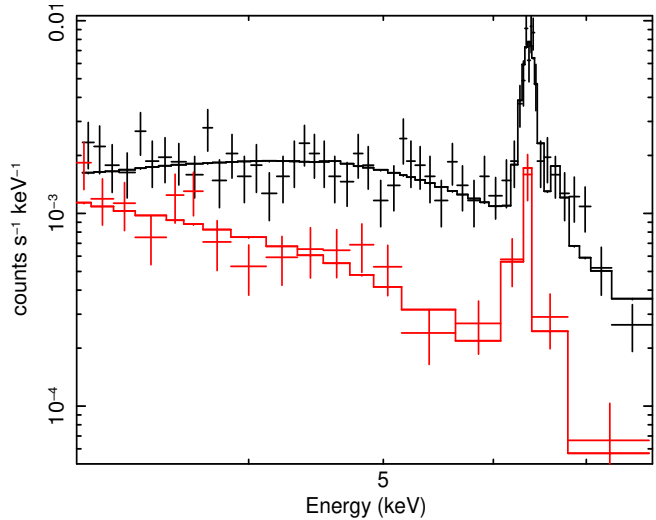

**Fig. 1** – 3-8 keV spectra from two different extraction regions with best-fit power-law + line models. The upper (black) spectrum is from the 1.5" nuclear circle, showing a flat continuum with a 6.4 keV Fe Kα line. The lower (red) spectrum is from the 1.5"-8" radii annulus, showing a steeper continuum and the prominent Fe Kα line.

This annular spectrum has a steep continuum and a strong Fe Kα line. The power-law + line model fit returns $\Gamma$ = 1.74(-0.41, +0.43) and EW = 2.10(-0.50, +0.52) keV. The 3-8 keV luminosity corresponding to this emission is ∼4.4 x $10^{39}$ erg s$^{-1}$, ∼24% of the entire luminosity in the same band from the 8" circle. In contrast, fitting the 1.5" nuclear circle (Fig. 1, black) using the same simple model gives $\Gamma$ = -0.36(-0.21, +0.19) and EW = 0.96±0.13. While the nuclear spectrum has far more counts and requires a more complex fit (Levenson et al. 2006, and Paper II in preparation,) this simple comparison makes the differences and relative strengths of the two components clear.

Fig. 2 shows the images derived from the data in the 3-6 keV (flat continuum) and 6-7 keV (Fe Kα) bands, binned in the native ACIS instrumental pixel (0.492''). Both bands show significant extended emission following the direction of the optical lines and radio emission (Falcke et al. 1998). We produced an azimuthal profile of the 3-6 keV emission, with angular bins of 5 degrees. The profile so obtained was fitted with two gaussians plus a constant to account for the background level. Based on the gaussian centers and widths, we selected the SE cone between 101 and 173 degrees (measured counter-clockwise from North) and the NW cone between 285 and 349 degrees.

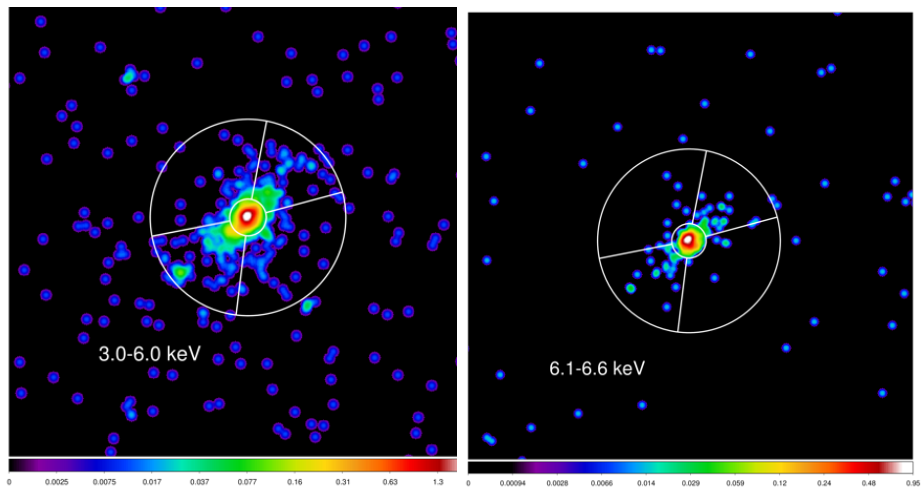

**Fig. 2** – Emission in the 3-6 keV continuum (Left), and the 6-7 keV (Fe Kα) band (Right). The image is in log intensity scale. The cone sectors, with 8'' outer radius (including the bulk of the extended emission) are also shown. The cone opening angles are 72 for the SE cone and 64 for the NW cone. The two wider sectors we refer to as the cross-cone direction. The inner circle is at a radius of 1.5''.

To quantify the magnitude and significance of the extended components in different energy bands, we have compared a set of radial profiles from those energy bands, with a set of *Chandra* Point Source Functions (PSF) models generated with CHART[3] and MARX[4] for the emission centroid position and the same energy bands. Off-nuclear point sources in the image were in all cases subtracted from the radial profiles, but their count contribution is in any case small (<3%). We normalized the PSFs to the counts within a circle of 0.5'' radius and, to further avoid contamination from a central component, we also excluded from the comparison the central 1.5'' radius circle, which is dominated by the central emission, where centroid placement may critically affect the results. Table 1 lists the excess counts above the PSF in various energy bands, together with the Poisson statistical error in each case, for counts extracted from an annulus of 1.5''- 8'' radii, and from the cone and cross-cone directions (as defined in figure 2).

---

[3] URL: http://cxc.harvard.edu/ciao/PSFs/chart2/
[4] URL: http://cxc.harvard.edu/ciao/threads/marx/

Table 1.
Excess counts over PSF from regions outside the central 1.5'' radius circle

| Energy (keV) | 8'' radius circle | | Cone Sectors | | Cross-Cone Sectors | |
| --- | --- | --- | --- | --- | --- | --- |
| | Excess | Err. | Excess | Err | Excess | Err. |
| 3-6 | 249.7 | 18.9 | 192.0 | 15.2 | 57.7 | 11.2 |
| 3-4 | 146.3 | 13.5 | 114.1 | 11.3 | 32.2 | 7.4 |
| 4-5 | 73.1 | 10.5 | 58.7 | 8.5 | 14.4 | 6.1 |
| 5-6 | 30.0 | 8.1 | 19.0 | 5.7 | 10.9 | 5.8 |
| 6-7 | 48.9 | 10.1 | 45.7 | 8.1 | 3.1 | 6.0 |
| 6.1-6.6 | 33.5 | 8.1 | 34.0 | 6.8 | -0.5 | 4.5 |
| 7-8 | -5.0 | 6.5 | 1.6 | 4.4 | -6.6 | 4.7 |

The 3-6 keV image (corresponding to figure 2 Left), contains 1070±33 net source counts within an 8'' radius circle, after excluding the non-nuclear point sources in the field. As can be seen from Table 1, there are ~250 counts in excess of what could originate from a nuclear point source, and the majority of this excess is in the direction of the optical line and radio emission (which we call the 'cone'; Fig. 2). Significant extended emission is detected in each of the 1 keV wide bands up to 6 keV, although the number of counts decreases with increasing energy, as also indicated by the steeper spectrum in the 1.5''- 8'' radii annulus (Fig. 1, Right). In the 6-7 keV band the counts increase because of the prominent 6.4 keV Fe K$\alpha$ line. In the 6.1-6.6 keV band, dominated by the Fe-K$\alpha$ line, there is a ~34 count excess. Since at most 8 of these counts may be due to continuum, the Fe K$\alpha$ line is extended at >3$\sigma$ significance. As for the 3-6 keV continuum this excess is mostly along the cone direction. Instead, we do not detect extended emission from the 7-8 keV continuum band. Fig. 3 shows the radial profiles of the 3-6 keV (Left) and 6.1-6.6 keV (Right) bands, including also the data within the central 1.5'' radius.

To investigate the two-dimensional morphology of the extended emission, we processed the 3-6 keV data, spatially rebinned with a resolution 1/8 of the instrumental pixel, with the EMC2 PSF deconvolution algorithm (200 iterations; Esch et al. 2004; Karovska et al. 2005, 2007; Wang et al. 2014). To enhance the image visually, the deconvolved image was then adaptively smoothed with the CIAO tool *csmooth*[5], with a Gaussian smoothing kernel at maximum smoothing scale of 2 image pixels. The resulting image (Fig. 4, Top) shows an elongated (~2 kpc diameter) low surface brightness region, within which is embedded the nucleus and a bright, jet-like, ~500 pc long extension to the SE, which is spatially correlated with the optical emission line and radio emission (Fig.4, Bottom; Ulvestad & Wilson 1989; Falcke et al 1998). The contours superimposed on the bottom left panel of Fig. 4 are Hα + continuum, from the archival HST 800s exposure of WF/PC2 with filter F658N (PI: Wilson), those on the bottom right panel are from the radio 6cm. VLA image (Observer: Ulvestad; retrieved from NED).

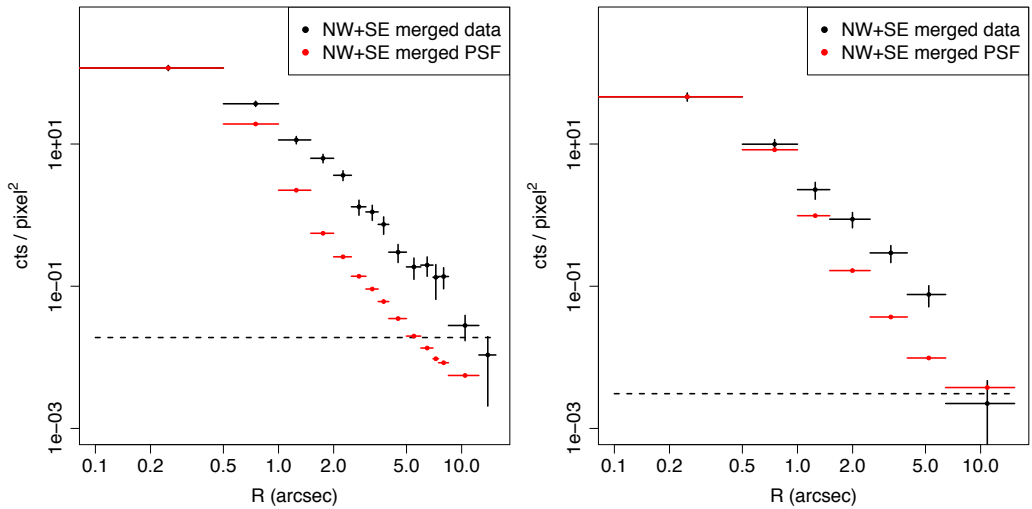

**Fig. 3** – Radial profiles of the emission in the cone sectors: Left, is the 3-6KeV continuum. Right, is the 6.1-6.6 Fe Kα band. 1/8 pixel re-binning was used. The NW and SE cones were merged radially, compared with the radial profiles of the PSF, normalized to the source image in the central 0.5'' circle. The bin size (shown) was chosen to contain a minimum of 10 counts. Errors are 1σ. The profiles are background subtracted. The background level is shown by the dashed horizontal line.

## 3. Discussion

In the CT AGN ESO428-G014 the hard continuum and Fe Kα emission are extended to large scales that reach an extent of ~2 kpc (~1 kpc radius **= 9''** radius; Figs. 3, 4; Table 1). Since ~24% of the 3-8 keV hard emission of ESO428-G014 arises outside the central 1.5'' radius region (and ~33% outside the central 1'' radius), we may need to revise the picture which is at the base of our interpretation of CT AGN spectra.

---
[5] URL: http://cxc.harvard.edu/ciao/ahelp/csmooth.html

The number of counts in the 1.5''-8'' spectrum implies a luminosity in the 3-8 keV band of ~4.4 × $10^{39}$ erg s$^{-1}$. Using established scaling relations (e.g., Lehmer et al 2010), this luminosity would be consistent with the total output of a population of X-ray binaries in a galaxy as luminous as ESO428-G014 ($L_V$~3 × $10^{10}$ $L_{Sun}$). However, it is spatially concentrated in a very small area of the galaxy, while X-ray binaries tend to be distributed over the entire galaxy stellar body (Fabbiano 2006). More important, the large Fe Kα EW = 2.10(-0.50, +0.52) keV in this extended emission (Section 2) excludes an origin due to unresolved X-ray binary emission (e.g., Barr et al. 1985). These large-scale components follow the direction of the extended soft X-ray (Levenson et al 2006) and optical line emission (Fig. 4, bottom left), and of the radio jet (Fig.4, bottom right, Ulvestad & Wilson 1989; Falcke et al 1998), suggesting that this emission is from phenomena connected with the AGN.

The differences between NGC4945, where the hard continuum and the Fe K emission are confined to a flattened 'torus' ~200 × 100 pc region perpendicular to main axis of the ionization cone (Marinucci et al 2012), and ESO428-G014, where a >2 kpc diameter extended component follows the ionization region, demonstrate that the continuum and fluorescent iron emission in CT AGNs are more complex than generally assumed. *Chandra* observations of the Circinus galaxy (Arevalo et al 2014) and NGC1068 (Bauer et al 2015) also show that both toroidal and more extended components are possible in the same AGN. In both cases, these extended components are on physical scales smaller that what we detect in ESO428-G014. In Circinus, both toroidal and non-toroidal components extend over ~30'' (~600 pc at a distance of 4.2 Mpc). In NGC1068, the *Chandra* data suggest extent of the Fe K line with ~30% of the emission on scales >140 pc.

The steeper spectrum of the annular 1.5''-8'' radii region (Γ~2), compared with the flat overall (fig. 1) and nuclear (see Levenson et al 2006) hard continuum, is different from the flat continuum seen in the 'torus' of NGC4595. This steep spectrum is typical of unobscured AGNs (e.g. Just et al., 2007), and so suggests a scattered contribution from the intrinsic Seyfert emission. In the standard Unified Scheme there is no screen in the cone direction, allowing the central source spectrum to escape unattenuated. The steep spectrum and the relatively low ionization parameters in these clouds far away from the nucleus are consistent with the large ~2 keV Fe Kα E. W. observed (Garcia et al 2013).

A contribution from line emission in the 3-4 keV band from photoionized emission may also result in a steepening of the spectrum. The brightest lines between 3 and 6 keV are Ar XVII (3.107 keV), Ar XVIII (3.321 keV), Ca XIX (3.876 keV), and Ca XX (4.106 keV). However, they should not be strong for gas with normal abundance ratios. It would take some fine-tuning to avoid getting Si and S emission an order of magnitude stronger than observed (Paper II, in preparation).

Levenson et al. (2006) find an intrinsic nuclear luminosity, L(2-10)=(3-7.6)×$10^{42}$ erg s$^{-1}$. The bi-cone has an observed, projected, half-opening angle of ~30° and hence roughly 13% of the unobscured central power is radiated in the direction of the two cones, or ~5×$10^{41}$ erg s$^{-1}$.

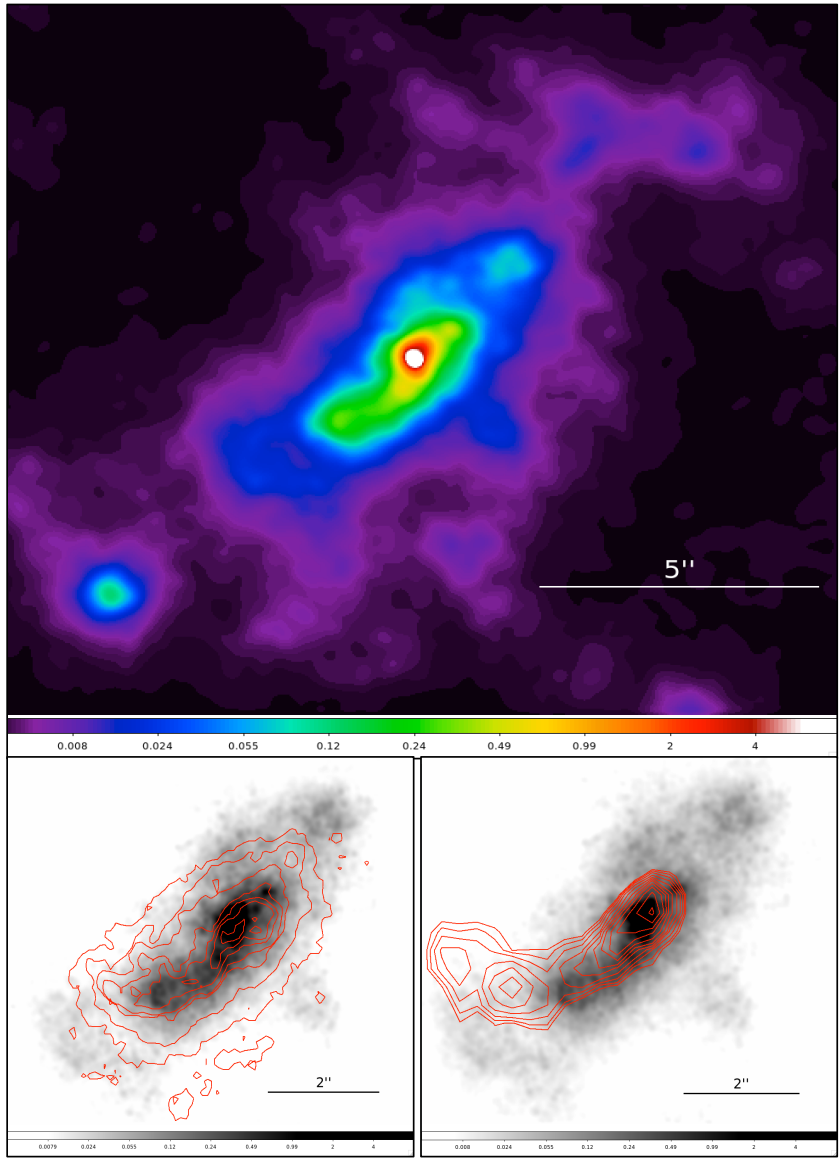

**Fig. 4** – Top: 3-6 keV 1/8 pix. deconvolved and adaptively smoothed image (see text). The point-like source to the SE is not connected with the extended component. The box is ~1.7 kpc side. Bottom Left: Zoom-in of the central higher intensity region, with Hα (+continuum) contours (see text). Bottom Right: same as bottom left, but with radio 6 cm contours (Ulvestad & Wilson 1989). The linear scale is 112pc/arcsec.

The true opening angle may be larger (Fischer et al. 2013). For an opening angle of 42° the fraction of the central power becomes 25%, giving an idea of the uncertainties in this calculation. We observe ~0.5% – 1% of this power in extended hard X-ray emission. So for a uniform medium to Thomson scatter this fraction of the intrinsic luminosity within the cones needs a total of $N_H \sim 3.5 \times 10^{21}$ cm$^{-2}$ of ionized gas in all directions of the bi-cone solid angle. Over 1 kpc the required column density corresponds to a physical density of ~1 cm$^{-3}$, a reasonable ISM value. At 600 pc radius, within which the bulk of the extended hard X-ray continuum is seen, the cone will have a half-width of ~300 pc, for a 30° half-opening angle. If what we observe is not the filled cone in projection, but the intersection of the cone with the galactic disk, with a typical 100 pc scale height, the scattering medium will be confined to a section of the cone that is still ~600 pc in radius, but only 100 pc in thickness. In this case scattering by higher density molecular clouds in the ISM of the galaxy disk will be required. Moreover, as we are observing the source at a large angle to the scattering medium, the angular dependence of the scattering becomes important. The photons that are scattered to our lines of sight (~90 deg), will then be a smaller fraction of the entire scattered flux, requiring a higher column density than in our simple geometric approximation. The scattering column density, though, cannot be larger than ~$7 \times 10^{22}$ cm$^{-2}$, unless the scattering gas is highly ionized, without producing a low energy photoelectric cut-off in the scattered spectrum. We do not observe such a cut-off in the 3-8 keV spectrum. Some of the scattering clouds may be as Compton thick as those in the obscuring torus. These could be normal ISM clouds, as a Compton thick molecular cloud ('the brick') has been reported near our own Galactic Center (Longmore et al. 2012).

As electron scattering is wavelength independent, the hard X-rays may be scattered by the same gas that scatters the broad emission lines from otherwise hidden broad emission line regions (Miller et al. 1991), but extending to at larger radii. An electron scattering origin predicts strong polarization of the extended hard X-ray emission (Elvis and Lawrence 1988). Hard X-ray imaging then allows us to determine the shape of the "mirror" that gives us a view into otherwise obscured active nuclei.

An AGN wind may also directly lead to the formation of dust (Elvis et al. 2002). The widespread presence of IR emission in local AGNs extending preferentially along the axis of the [OIII] emission (Asmus et al 2016) is consistent with a population of dense dusty clouds. There may also be a population of optically thick clouds with an optically thin 'skin'. The same photoionized material that produces the low energy line emission radiation (see, e.g. Kraemer et al. 2000, Wang et al. 2011b) could be responsible for the extended hard X-ray and Fe-K emission. In this case, we should find, on average, correspondence not only between [OIII] and soft X-ray features (see Wang et al 2011a,b), but also with harder continuum. This may occur in ESO428-G014, since the soft, line-dominated, X-ray emission extends in the same region as the hard continuum (Levenson et al 2006; Paper II, in preparation).

The magnetic fields in the radio jet of ESO428-G014 (Ulvestad & Wilson 1989) may instead accelerate the charged particles of the entrapped plasma, with consequent hard X-ray emission (e.g., Petrosian 2001), resulting in the good correspondence between radio and hard X-ray features (Fig. 4, Bottom Right). Deep high-resolution *Chandra* ACIS observations of nearby Seyfert 2s have shown that radio jets may also produce collisional excitation of encountered clouds (see

NGC4151 and MKN573, Wang et al. 2011a,b; Paggi et al. 2012). Interestingly, in ESO428-G014 the X-ray surface brightness increases at the bend in the jet, suggesting an interaction (Fig. 4, Bottom Right). If the shock velocities are >1000 km s$^{-1}$, hard X-ray emission will result. For shocks up to ~2500 km s$^{-1}$, a prominent Fe XXV 6.7 keV line is expected. This may occur in the CT AGN merger NGC6240 (Wang et al. 2014), but in the extended emission of ESO428-G014 we only detect neutral 6.4 keV Fe K$\alpha$; the data do not allow us to constrain the temperature of an additional thermal component. Ultrafast nuclear outflows (UFOs, $v_{out}$ > 10,000 km s$^{-1}$, see Tombesi et al. 2013) may interact with and shock the ISM, producing a hard thermal continuum, at temperatures too high to produce Fe XXV. So in this case, a hard continuum would be expected but not extended Fe emission, which is contrary to our data. A better determined continuum slope would be diagnostic.

## 4. Conclusions

The deep 154 ks cumulative *Chandra* ACIS-S exposure on the CT Seyfert galaxy ESO428-G014 has led to the discovery of kpc-scale diffuse emission in the 3-6 keV band and in the Fe K$\alpha$ line. This extended hard component contains at least ~24% of the 3-8 keV emission. This is the first reported instance of hard and Fe K$\alpha$ emission extended on kpc size scale. This discovery suggests that we may need to revise the picture at the base of our interpretation of CT AGN spectra.

In both line and continuum, the emission follows the direction of the extended optical line emission (ionization cone) and radio jet. A uniform scattering medium of density 2.5 cm$^{-3}$ would produce this luminosity in a 1 kpc path length in the bi-cones. Alternatively, higher column density molecular clouds in the disk of ESO428-G014, intercepted by the bi-cones, may be responsible for these components. If the hard X-rays are scattered by the same gas that scatters the broad emission lines, the hard X-ray emission should be strongly polarized. Hard X-ray imaging then lets us study the "mirror" that provides this indirect view of otherwise obscured AGN.

The steeper spectrum of the hard continuum outside of the central 1.5'' nuclear region suggests that this emission may be due to a contribution of intrinsic Seyfert emission. The continuum may also be enhanced by the acceleration of charged particle in the radio jet. Ultrafast nuclear outflows cannot explain the extended Fe K$\alpha$ emission.

We will pursue a detailed spectral analysis and comparison with the softer X-ray emission line emission, and with optical and radio data in future work (Paper II). Higher signal-to-noise X-ray spectra at arc-second resolution to accurately determine the spectral parameters could distinguish between models. These results demonstrate the importance of joint high spatial and spectral resolution for X-ray observatories.

We thank Aneta Siemiginowska for discussions on radio jets and feedback. We retrieved data from the NASA-IPAC Extragalactic Database (NED), the *Hubble Space Telescope* Archive, and the *Chandra* Data Archive. For the data analysis we used the *CIAO toolbox*, *Sherpa* and *DS9*, developed by the Chandra X-ray Center (CXC); and *XSPEC* developed by the HEASARC at NASA-GSFC. This work was partially supported by the *Chandra* Guest Observer program grant GO5-16090X (PI: Fabbiano), and by NASA contract NAS8-03060 (CXC).